\documentclass[11pt ]{article}
\usepackage[dvips]{epsfig}
\newcommand{\ee}{$e^+e^-\: $}

\newcommand{\qq}{$q\bar{q}\:\;$}

\newcommand{\gaga}{$\gamma \gamma\:\;$}

\newcommand{\hZg}{$H \rightarrow Z \gamma  \:$}
\newcommand{\noi}{\noindent}
\newcommand{\BS}{\bigskip}
\begin{document}

\pagestyle{empty}

\noi DESY 03-017    \hspace{12.0cm} LC-PHSM-2003-004

\BS

\noi February 2003

\vspace{2.5cm}
\begin{center}
\LARGE{\bf\boldmath Measurement of the $H Z \gamma$ Coupling \\
                    at the Future Linear \ee Collider \unboldmath}
\end{center}

\vspace{1.0cm}
\large
\begin{center}
M. Dubinin$^1$, H.J. Schreiber$^2$ and A. Vologdin$^1$
\end{center}

\vspace{0.3cm}
\bigskip \bigskip  
\begin{center}
\small
$^1$ Institute of Nuclear Physics, Moscow State University, \\
119 992 Moscow, Russia \\ [2mm]

$^2$ DESY Zeuthen, 15735 Zeuthen, FRG \\ [2mm]

\normalsize
\end{center}

\vspace{1.0cm}
\pagestyle{plain}
\pagenumbering{arabic}

%
\section*{Abstract}

We examine the prospects for measuring the $H Z \gamma$ coupling of a
Standard Model-like Higgs boson with a mass between 120 and 160 GeV
at the future TESLA linear \ee collider, assuming an integrated luminosity 
of 1 ab$^{-1}$ and a center-of-mass energy of 500 GeV. We consider
the Higgs boson produced in association with $\nu_e \bar{\nu_e}$
via the $W W$ fusion reaction $e^+e^- \rightarrow \nu_e \bar{\nu_e} H$,
followed by the rare decay into a $Z$ boson and a photon, $H \rightarrow
Z \gamma$. Accounting for all main background contributions, different
selection procedures are discussed. Uncertainties on the \hZg branching 
fraction of approximately 48\% (27\%, 44\%) can be achieved in unpolarised
\ee collisions for $M_H$ = 120 (140, 160) GeV. With appropriate initial
state polarisations $\Delta$BF(\hZg)/BF(\hZg), or the precisions on the
\hZg partial width, can be improved to 29\% (17\%, 27\%) and provide valuable
information on the $H Z \gamma$ coupling.
We regard our results as a convincing confirmation of the great potential
of a linear collider to access and reliably measure important parameters 
of the Higgs boson despite initially overwhelming background with 
final state signature similar to the signal events.

\newpage

\section {Introduction}

 Following the discovery of the Higgs boson, one of the main tasks
 of a future linear
 \ee collider will be precise model-independent measurements
 of its fundamental
 couplings to fermions and bosons and its total
 width \cite{LC1}. In particular, the determination of the couplings
of the Higgs boson to the other fundamental particles will be
a crucial test of the nature of the Higgs particle. In this respect
future linear colliders will play a major role.
Different colliding options with different beam polarisations
combined with adjustable center-of-mass energies in a wide range
and the clean enviroment in these machines will allow for rather
precise determinations of these couplings.

A lot of detailed studies of the Standard Model (SM) Higgs couplings
to fermions and $W$ and $Z$ bosons can be found in the literature
\cite{couplings}. These studies demonstrate the ability of a
linear collider to access these couplings with precision
of a few percent.
Also the trilinear Higgs self-coupling
in the double Higgs production processes \ee $\rightarrow Z H H$
and \ee $\rightarrow \nu_e \bar{\nu_e} H H$
\cite{tricoupl} are within the possibilities of experimental
verification, although with substantially lower precision.

Another set of important Higgs boson couplings is represented
by the effective vertices $Hgg$,
$H\gamma\gamma$ and $HZ\gamma$. These couplings do not occur
at the tree-level but are induced by loop diagrams \cite{loop}.
Since the Higgs interaction is proportional to particle masses,
loop contributions of massive fermions do not decouple,
and these vertices
could therefore serve to count the number of particles which couple
to the Higgs boson. The $Hgg$ vertex can be accessed through the
$H \rightarrow gg$ decay in \ee collisions \cite{Hgg} or in the
fusion reaction $gg \rightarrow H$ at the LHC \cite{HggLHC}. The
$H\gamma\gamma$ coupling can be determined 
either in \ee and LHC-$pp$ interactions
when the Higgs decays into two photons \cite{shanidze},\cite{gagaLHC}
or directly by means of the Compton back-scattering \gaga fusion
process \gaga $\rightarrow H\rightarrow X$, with probably
the best precision \cite{gammagamma}.

\vspace{3mm}
In this study we explore the potential of a linear \ee collider to
measure the $HZ\gamma$ coupling through the rare \hZg decay, for
masses of the Higgs particle in the range 120 to 160 GeV.
Precise electroweak data provide the existence of a light Higgs
boson  with a mass below 193 GeV with 95\% confidence level \cite{La},
with a preference
for $M_H$ close to 120 GeV. There are also hints of a signal
in the direct search in \ee $\rightarrow HZ$ at LEP2,
with a lower mass limit of $M_H \ge$ 114.1 GeV at 95\% CL \cite{LEP}.

\vspace{3mm}
The reaction which will be used to explore the branching
fraction BF(\hZg) is
\begin{equation}
       e^+e^- \rightarrow \nu_e \bar{\nu_e} Z \gamma \hspace{2mm},
\end{equation}
assuming a Higgs boson mass $M_H$ = 120, 140 and 160 GeV,
$\sqrt{s}$ = 500 GeV and an integrated luminosity 
of ${\cal L}$ = 1 ab$^{-1}$.

The statistical precision for
BF(\hZg) is mainly determined by
$\sqrt{S+B}/S$, where $S$ and $B$ are respectively the
number of signal and background events 
within a small interval of the $Z\gamma$
invariant mass, centered around $M_H$. Hence, evaluation
of all relevant signal and background processes and
optimization of selection procedures are mandatory,
taking into account acceptances and resolutions of a linear collider
detector.

Our analysis is, to our knowledge, the first on the study
of BF(\hZg) at any (future) collider.
It includes the complete irreducible background
and all main reducible background contributions expected.
Since the SM branching fraction \hZg is very small
and there is in particular
large irreducible background, the most important hadronic Z
decays, $Z \rightarrow q\bar{q}$, are
accounted for in this analysis. Although
leptonic (e, $\mu$) $Z$ decays provide
a clean final state signature and best $Z$ boson recognition,
their rates are however too small to include them at this stage of
the analysis or to allow for an independent approach.
$Z \rightarrow \nu \bar{\nu}$ decays are not considered
since they prevent $Z$ boson reconstruction in reaction (1).

\vspace{3mm}
 The paper is organized as follows. In Section~2 we discuss
 simulation of the Higgs signal and background events and
 their detector response. In Section~3 we present
 the results for unpolarised BF(\hZg) measurements,
 using different selection procedures.
 In Section~4 improvements to the \hZg branching 
fraction measurements are discussed
 when e.g. beam polarisation is accounted for in
signal and background events. Also, expectations on BF(\hZg) from the
Higgs-strahlung process and possible systematic errors and
the effect of overlap with $\gamma \gamma \rightarrow hadrons$
are reviewed. Section~5 summarizes the conclusions.

\begin{center}
\begin{figure}[t]
\begin{minipage}[b]{18cm}
\vspace{-0.2cm}
\epsfig{file=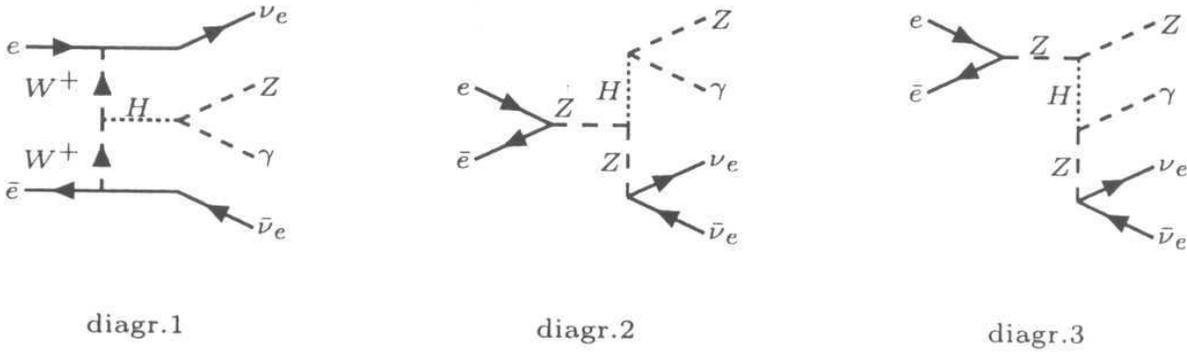,height=5cm,width=16cm}
\caption{Signal diagrams for reactions (2) and (3), with
 $Z \rightarrow \nu_e \bar{\nu_e}$ and \hZg decays.}
\end{minipage}
\end{figure} 
\end{center}

\vspace{-1.1cm}
\section {Event generation}

 In $e^+e^-$ collisions the Standard Model Higgs boson is predominantly
produced by two different processes, the Higgs-strahlung process
\begin{equation}
      e^+e^- \rightarrow Z H 
\end{equation}
and the weak boson ($WW$ and $ZZ$) fusion reactions \\
\begin{equation}
      e^+e^- \rightarrow \nu_e \bar{\nu_e} H   
\end{equation}
\begin{equation}
      e^+e^- \rightarrow e^+ e^- H
\end{equation}
   
The $ZZ$ fusion process (4) is strongly suppressed with
respect to the $WW$ fusion process (3) by about a factor of 10,
rather independent of $\sqrt{s}$.
Therefore, the production channel (4) is ignored in this study.
The SM tree-level diagrams contributing to the signal reactions (2)
and (3), including the $Z \rightarrow \nu_e \bar{\nu_e}$ and \hZg decays,
are shown in Fig.~1.

\vspace{3.0mm}
In the $production \times decay$ approximation the processes (2) and (3)
are factorizable parts of the Higgs signal diagrams for the
2-to-4 body reaction (1) with electron neutrinos in the final state.
In other words, the amplitude squared of diagram 1 in Fig.~1
integrated over the phase space gives
$\sigma$(\ee $\rightarrow \nu_e \bar{\nu_e} H)
\cdot BF(H \rightarrow Z \gamma)$
and the amplitudes squared of diagrams 2 and 3 give
$\sigma(e^+e^- \rightarrow Z H) \cdot BF(Z \rightarrow \nu_e \bar{\nu_e})
\cdot BF(H \rightarrow Z \gamma)$.
Thus, to be most general in our analysis, events
of reaction (1) were generated for the complete set of tree-level
diagrams (see Fig.~2 for the contributing background diagrams)
by means of the program package CompHEP \cite{CompHEP}, including
initial state bremsstrahlung and beamstrahlung for the TESLA linear
collider option \cite{bms}. 

\begin{center}
\begin{figure}
\vspace{-3.0cm}
\includegraphics[width=0.95\textwidth,height=.95\textheight]
  {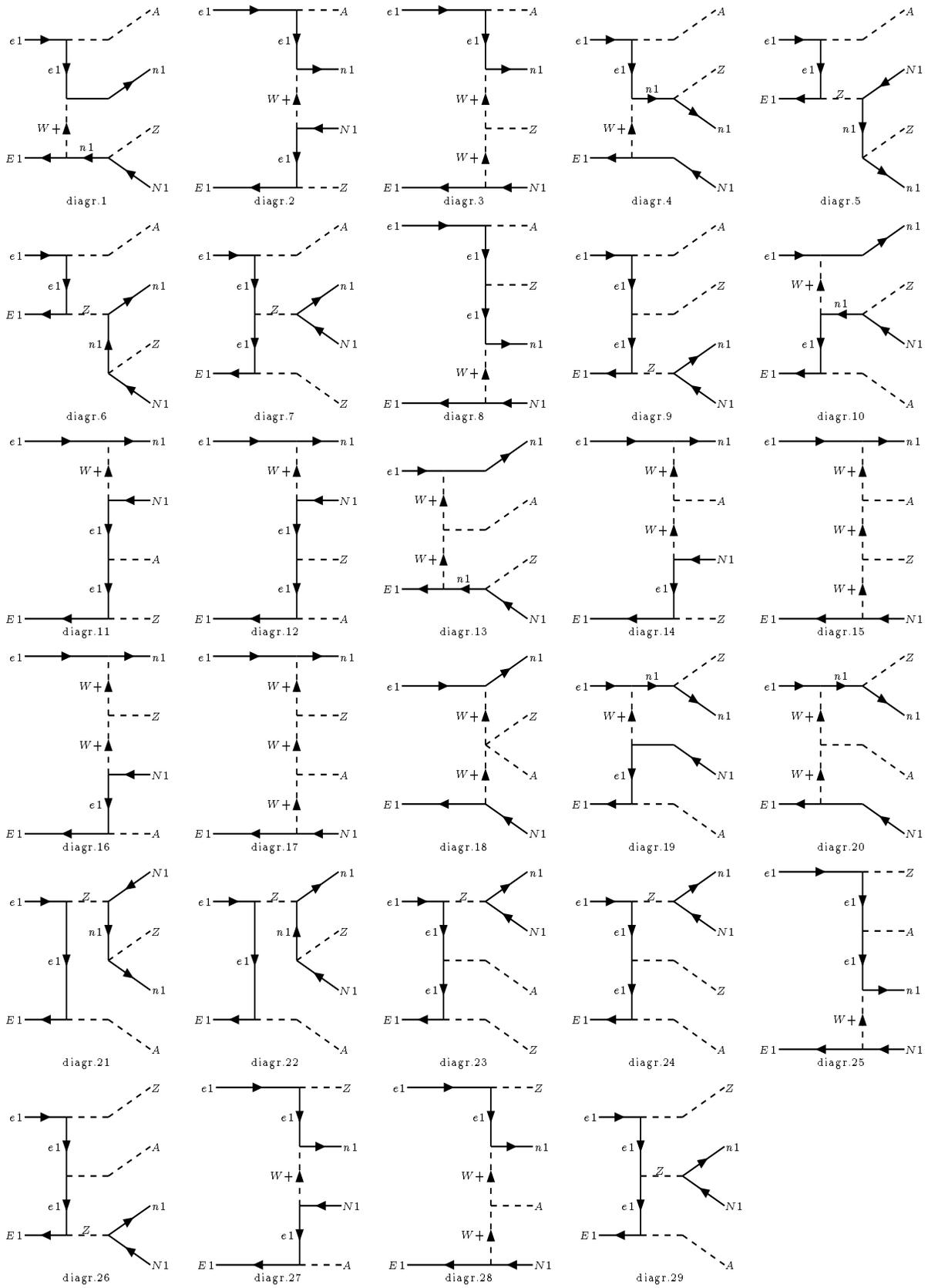}
\caption{Background diagrams for the reaction
         $e^+e^-\rightarrow\nu_e\bar{\nu_e} Z \gamma$.}
\end{figure}
\end{center} 

\vspace{-1.0cm}
The present version of CompHEP performs analytic calculations
of the matrix element squared, generates an optimized Fortran code
and generates a flow of events. In addition, it provides for the user
an appropriate kinematical scheme for the integration
over the four-body phase space. The basic input
parameters are taken from the report of the Particle Data Group
\cite{PDG} or are as listed here: $m_{b}$ = 4.3 GeV, $\alpha_{EW}$ =1/128,
$M_{Z}$ = 91.19 GeV, $sin^2\theta_W$ = 0.23 and $\Gamma_Z$ = 2.50 GeV.
The CompHEP-PYTHIA interface package \cite{interface} was used to
simulate the $\nu_e\bar{\nu_e}Z\gamma \rightarrow 
\nu_e\bar{\nu_e}q\bar{q}\gamma$ signature.
In this way, Higgs boson production
and the complete irreducible background as well as possible
interferences are taken into account.

For unstable particles, Breit-Wigner formulae have been used for the
s-channel propagators. The Higgs boson width and
the \hZg branching fraction were imported from the program
package HDECAY \cite{HDECAY}. BF(\hZg) depends on the Higgs mass and
is largest near $M_H$ = 144 GeV. Some values
of this branching fraction, the total Higgs width and the
$HZ\gamma$ effective coupling constant relevant for our study
are summarized in Table 1.

\begin{center}
\vspace{0.4cm}
\begin{tabular}{c|c|c|c} \hline \hline
M$_H$, GeV  &  BF(\hZg)  &  $\Gamma_{tot}$, GeV  &  $\lambda_{HZ\gamma}$
 \\ \hline\hline 
 120 & 1.1$\cdot10^{-3}$ & 3.6$\cdot10^{-3}$ & 5.4$\cdot10^{-5}$ \\ \hline
 140 & 2.5$\cdot10^{-3}$ & 8.1$\cdot10^{-3}$ & 6.2$\cdot10^{-5}$ \\ \hline
 160 & 1.2$\cdot10^{-3}$ & 8.1$\cdot10^{-2}$ & 8.8$\cdot10^{-5}$ \\ \hline
\end{tabular}
\end{center}
 Table~1: Branching fractions, total widths and effective coupling
constants of the SM Higgs boson for $M_H$ = 120-160 GeV. \\

As discussed in ref. \cite{our}, a favored signal to
background situation is expected for the Higgs $WW$ fusion reaction (3)
at e.g. $\sqrt{s}$ = 500 GeV,  and we will consider only
this case in the following.

Due to the small cross-section expected for the signal reaction
$ e^+e^- \rightarrow \nu_e \bar{\nu_e} H \rightarrow \nu_e \bar{\nu_e}
 Z \gamma$, diagram 1 in Fig.~1,
we only rely on events with the most important
$Z\rightarrow $ \qq decays. Therefore, events of process (1)
are characterized by two hadronic jets
originating from the $Z$ boson, together with an energetic photon
and large missing energy due to the two final state neutrinos.
The invariant mass of the $Z$ and the photon should equal
$M_H$.

\vspace{3mm}
The irreducible background expected from reaction (1)
(the diagrams in Fig.~2) was accounted for
at the same level as the signal events. Contributions from Z decays into
$\mu$- and $\tau$-neutrinos which would occur from diagrams 7, 9, 23, 24,
26 and 29 were effectivelly removed by an appropriate missing
mass cut (section 3). Possible contributions from diagrams 5, 6, 21 and 22
with $n_1=N_1=\nu_{\mu}$ or $\nu_{\tau}$ were also calculated and found
to be negligible due to large off-shell $n_1 \rightarrow Z N_1$ decay with
the Z boson close to its nominal on-shell mass value.
The surviving $\nu_{\mu}$ and $\nu_{\tau}$ background rates were
found to be smaller than 1\% of the total irreducible background.
An important part of the background was found to arise from the
W-exchange diagrams, but significant contributions were
also found to be due to the
single bremsstrahlung production process $e^+e^- \rightarrow Z Z \gamma
\rightarrow \nu_e\bar{\nu_e} Z \gamma$.

\vspace{0.3cm}
As the cross-section for the irreducible
background is more than two orders of magnitude larger than the
signal cross-section, we first applied the following principal cuts
at the generation level, to both the signal and
background events:

\begin{itemize}
\item the photon energy should exceeds 10 GeV and
\item the polar angle of the photon should lie in the
  range 5 to 175 degrees.
\end{itemize}

After these criteria, practically all ($\sim$96\%) Higgs events
survive, while the overwhelming irreducible background was
substantially reduced. The cuts also largely avoid any
infrared and collinear singularities in the calculation of the
background amplitudes, as might be deduced from diagrams in Fig.~2.

\vspace{0.3cm}
Possible reducible backgrounds to
\ee$\rightarrow \nu_e \bar{\nu_e} H$ events which might mimic
the signal such as the large event rate reactions
\ee $\rightarrow W W (\gamma), e \nu W (\gamma)$,
$e^+ e^- (\gamma^*/Z) (\gamma), t \bar{t} (\gamma)$,
$W W Z (\gamma)$ and $Z Z Z (\gamma)$, with beamstrahlung,
initial state radiation, final state radiation and radiation
from the W boson itself,
were generated by either PYTHIA \cite{PYTHIA} or CompHEP \cite{CompHEP}.
The \ee $\rightarrow e \nu W (\gamma), e^+ e^- (\gamma^*/Z) (\gamma)$
events were obtained by $e-\gamma-e$ splitting and subsequent
$\gamma e \rightarrow \nu W$ respectively 
$\gamma e \rightarrow e (\gamma^*/Z)$
interactions by PYTHIA, with proper cross section normalizations.
Only those events were used for further analyses
if at least one final state photon exists
with principal cut properties.
It has been found that after detector
response simulation and enforcing the same selection procedures
as for the signal events (see below), $t \bar{t} (\gamma)$,
$W W Z (\gamma)$ and
$Z Z Z (\gamma)$ events were effectively discarded,
also if the two $W$ bosons from the top quarks and the prompt
produced $W$'s decay leptonically leading 
together with $Z\rightarrow$ \qq
to a topology similar to the signal topology.
Events from the \ee $\rightarrow W W (\gamma)$, $e \nu W (\gamma)$ and
$e^+ e^- (\gamma^*/Z)$ processes could however not be removed
to a negligible level.
Their contributions will be discussed below for each
selection procedure applied.

\vspace{3mm}
A further potential background is expected from the process
$e^+e^- \rightarrow q\bar{q}\gamma$, where initial state radiation and beamstrahlung
reduce the center-of-mass energy available close to the Higgs mass values.
Only events with center-of-mass energy below 200 GeV, a \qq system consistent with the
Z boson and a photon with large transverse energy were accepted and enforced
to the selection procedures. We found that radiative return events add some non-negligible
background to the final reconstructed $Z \gamma$ mass,
with some uncertainties due to the ISR model used.

\vspace{3mm}
Also possible contributions from Higgs-strahlung process (2) with
the dominant $H \rightarrow b \bar{b}$ and $WW^*$ decays were accounted for and found
to contribute with at most four eevents
thanks to our dedicated selection procedures.

\vspace{3mm}
All surviving reducible background events were included in the final
$Z \gamma$ invariant mass distributions and taken into account in
precision estimations for the \hZg branching fraction.

%
\section {\boldmath BF(\hZg) measurements at 500 GeV \unboldmath}

Based on the results of ref.\cite{our} that the
branching fraction BF(\hZg) should be preferably best measured
in the $WW$ fusion process (3) at a high-luminosity linear \ee collider,
we examine the prospects of measuring this quantity at  $\sqrt{s}$ = 500
GeV, for Higgs boson masses of 120, 140 and 160 GeV and an integrated
luminosity of 1 ab$^{-1}$. Since the $WW$ fusion cross-section rises
logarithmically with $\sqrt{s}$, large energies are
mandatory and the very small SM \hZg branching ratio
requires large accumulated luminosity.

\vspace{0.3cm}
The detector response for all generated signal and background
events was simulated with the parametrized detector simulation
program SIMDET$_{-}$v4 \cite{SIMDET} using parameters
as presented in the Technical Design Report \cite{TDR}.

\vspace{0.3cm}
Throughout this paper, it was demanded that each reconstructed
event involves more than three charged particles in the final state
and the total visible energy is less than 240 GeV with the transverse
component relative to the beam direction
below 210 GeV. Important for further analyses
is the requirement that the missing mass (caused by the two undetected
neutrinos) lies between 180 and 400 GeV. This cut ensures clean
elimination of the Higgs-strahlung process, $e^+e^- \rightarrow Z H
\rightarrow \nu_e \bar{\nu_e} Z \gamma$, with a
missing mass close to the $Z$ boson mass.

\vspace{0.3cm}
Large background event samples and tiny signal event rates
need to pursue different strategies for extracting signal events.
In order to account for the distinct properties of the final
state $\nu_e\bar{\nu_e}q\bar{q}\gamma$, we start
with a conventional method using
consecutive cuts on kinematical variables, while
more sophisticated selection procedures are followed to hopefully achieve
better signal-to-background event ratios and
hence smaller uncertainties on BF(\hZg) or the
$H Z \gamma$ coupling.

\subsection {\boldmath Event selection using consecutive cuts \unboldmath}

In a first attempt, a conventional method using simple consecutive
cuts was applied to isolate Higgs signal events.
We start to account for the
distinct properties of the photon from the Higgs decay, denoted as
$\gamma_H$, by demanding an energy greater than 20 GeV
with a transverse component of not less than 15 GeV.
In events where more than one photon candidate exists
the photon with largest transverse momentum was
selected as the Higgs decay candidate.
Thus, most of the bremsstrahlung photons with relatively
small energy at low polar angles are eliminated.
Furthermore, the $\gamma_H$ candidates should have
no particle in a cone of half-angle of $10^o$ around its direction,
i.e. they are demanded to be isolated.
Once a Higgs decay photon candidate had been found, it was removed
from the list of all final state particles, for which in turn, the
jet clustering algorithm of PYTHIA (subroutine PUCLUS) \cite{PYTHIA}
was enforced to isolate two jets. Each jet was required
to pass the following cuts:

\begin{itemize}
\item the number of particles is greater than 3;
\item the jet energy exceeds 8 GeV;
\item the angle between any two jets is larger than $20^o$;
\item the jet polar angle is within $|cos\Theta_{jet}| < 0.95$.
\end{itemize}

Thus, only well-measured and clearly separated jets were accepted.
The compatibility of the
dijet system with the $Z$ boson hadronic decay
was quantified by demanding that its invariant mass is within
84 to 105 GeV. The lower limit of 84 GeV was chosen
to reject part of the otherwise large $W W(\gamma)$
and $e \nu W (\gamma)$ background contributions. 
This requirement indicates
the importance of sufficient dijet invariant
mass resolution of the detector in order
to differentiate $Z \rightarrow$ \qq from $W\rightarrow q\bar{q}^{'}$ decays.
Then this dijet system was combined 
with $\gamma_H$ to establish the Higgs
particle in the $Z\gamma$ invariant mass. In order to enforce
an improved mass resolution of the $Z\gamma$ system,
a 1-constraint fit requiring $M_{jj} = M_Z$ was performed.

\vspace{0.3cm}
These selection criteria were not varied with the
Higgs masses considered.
It was found that changes of the cut parameters used
within reasonable limits had small or
negligible effects on the signal-to-background ratios.
Most of the remaining reducible background was due to the
\ee $\rightarrow W W (\gamma) \rightarrow l \nu q\bar{q}^{'}(\gamma)$ and
$e \nu W(\gamma) \rightarrow e \nu q\bar{q}^{'}(\gamma)$ channels
as well as radiative return events with an event
topology similar to the signal event topology.
We found that this background amounts to approximately 38\% of the
irreducible background for Higgs masses
of 140 and 160 GeV, while for $M_H$ = 120 GeV it was close to 16\%.
For the 120 GeV Higgs case, few Higgs-strahlung \ee $\rightarrow ZH$
events were found to survive.
Also some $e^+ e^- (\gamma^*/Z)(\gamma)$ events were retained above
$M_{Z\gamma}$ = 130 GeV. 

\vspace{0.3cm}
By following the strategy outlined above and assuming
${\cal L}$ = 1 ab$^{-1}$ of integrated luminosity, we obtain
the $Z\gamma$ invariant mass distribution for the surviving signal
and background events as shown in Fig.~3.
Higgs signals are evident on non-negligible background, with best 
significance for $M_H$ = 140 GeV. Clearly, the \hZg decay mode is not
favoured for Higgs boson searches but it allows to estimate the
branching fraction for Higgs masses close to 120 or 160 GeV.

\vspace{3.0mm}
Selection efficiencies for the signal and the irreducible background,
the number of signal $(S)$ and background $(B)$ events in the $Z\gamma$
mass range between 117 and 123 (137-143, 157-163) GeV,
the significances
$S/\sqrt{B}$ and the statistical precisions $\sqrt{S+B}/S$ obtained on
$\sigma$(\ee $\rightarrow \nu_e \bar{\nu_e} H)\cdot
BF(H \rightarrow Z \gamma)\cdot BF(Z \rightarrow q\bar{q})$
are presented in Table~2. The rather small signal selection
efficiencies obtained are mainly due to the difficulty to select
$\gamma_H$, the photon from the Higgs decay, out of all photons
in the final state. Stringent requirements on its energy and
transverse momentum, together with further cuts on event and jet
properties, were found to be neccessary in order to obtain
acceptable signal-to-background event rates.

\vspace{3.0mm}
Accounting for all $Z$ decays and the surviving reducible background
the relative precisions on the \hZg branching fraction
are then deduced after convolution with the uncertainty of the
inclusive $\nu_e \bar{\nu_e} H$ cross-section of few percent \cite{TDR}:
$\Delta$BF(\hZg)/BF(\hZg) = 65\% (28\%, 57\%)
for $M_H$ = 120 (140, 160) GeV,
assuming 1 ab$^{-1}$ integrated luminosity
at $\sqrt{s}$ = 500 GeV.


\begin{center}
\begin{figure}[t] 
\begin{minipage}[t]{18cm}
\epsfig{file=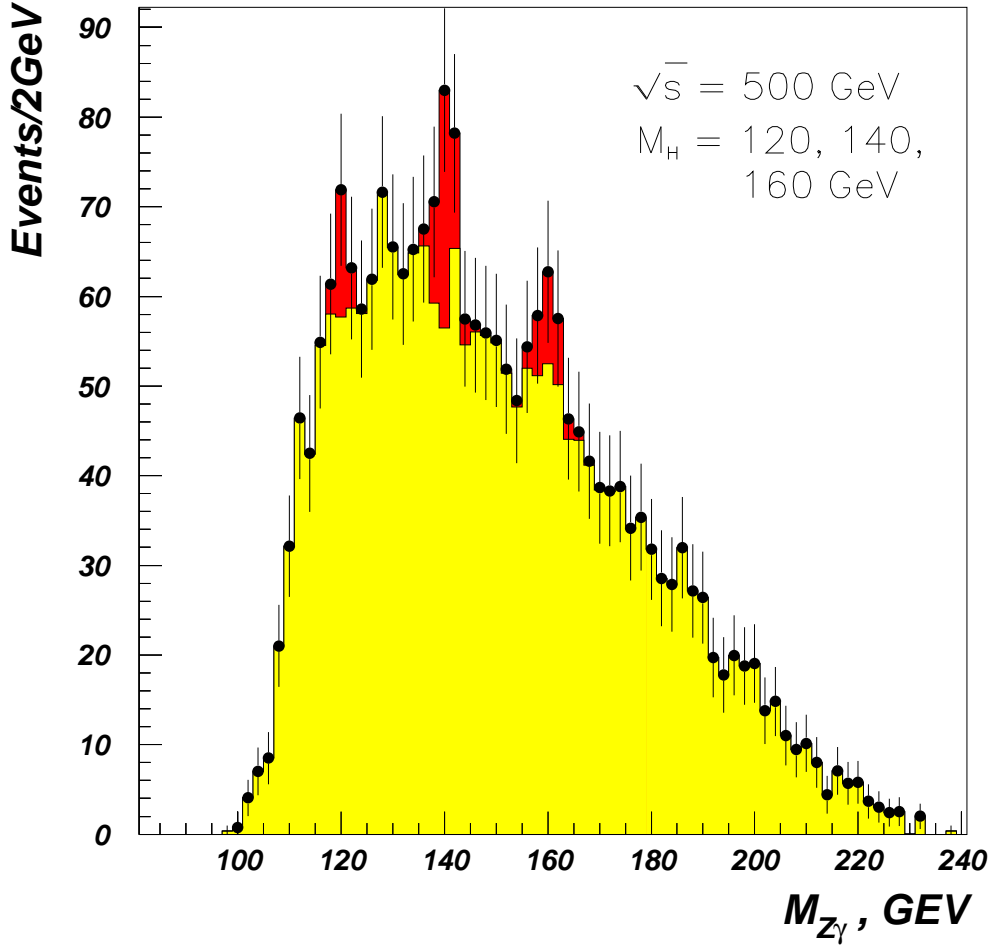,width=15cm}
\caption{ $M_{Z \gamma}$ invariant mass distribution for surviving
     signal and background events at $\sqrt{s}$ = 500 GeV,
     assuming 1 ab$^{-1}$ integrated luminosity.}
\vspace{0.5cm}
\end{minipage}
\end{figure}
\end{center}

\begin{center}
\begin{tabular}{c|c|c|c} \hline \hline
               & \multicolumn{3}{c}{ Signal \hspace{3mm}(Background)}
 \\ \hline
               & $M_H$ = 120 GeV   &  140 GeV  &  160 GeV
 \\ \hline\hline
Selection efficiency (\%)   & 23.9 (0.67) & 27.7 (0.52) & 24.9 (0.45) \\ \hline
Number of events/1 ab$^{-1}$     & 18 (149) & 45 (116) & 19 (99) \\ \hline
Significance $S/\sqrt{B}$        & 1.47  & 4.18  & 1.91 \\ \hline
Precision $\sqrt{S+B}/S$         & 0.72  & 0.28  & 0.57 \\ \hline
\end{tabular}
\end{center}
 Table~2: Selection efficiencies (see text) for both signal and
background $\nu_e\bar{\nu_e}q\bar{q}\gamma$ signature,
together with significances and precisions of
$\sigma$(\ee $\rightarrow \nu_e \bar{\nu_e} H)\cdot
BF(H \rightarrow Z \gamma)\cdot BF(Z \rightarrow q\bar{q})$
for $M_H$ = 120, 140 and 160 GeV, at the \ee linear collider with
$\sqrt{s}$ = 500 GeV. A total integrated luminosity of 1 ab$^{-1}$
is assumed. \\

\subsection {\boldmath Event selection using a jet finder \unboldmath}

The events generated by CompHEP or PYTHIA
including the principal cuts and
reconstructed by SIMDET were passed through a jet cluster
algorithm.  
The concern is that the jet finder is able to isolate the Higgs decay
photon, $\gamma_H$, from all other final state particles due to
its distinct properties, namely
the high transverse momentum and its relative
isolation from all other particles.
The remaining particles in the final state were then
clustered in two hadronic jets, with an invariant mass compatible with
$Z \rightarrow$ \qq decays, i.e. $M_{jj}$ was required to lie in
the range 84 to 105 GeV \footnote {For the few events with three
isolated hadronic jets it was demanded that their invariant mass
lies within the same limits.}.
We applied two cluster algorithms in order
to obtain some confidence on the jet finders and
to control systematic uncertainties in extracting signal and
background events. Subroutine PUCLUS of the PYTHIA package \cite{PYTHIA}
was enforced to isolate three or more jets, with one jet to be
consistent with a photon. The DURHAM algorithm \cite{Durham}
with $y_{cut}$ = 0.004 was applied to isolate events with at least
three jets, with one jet
required to pass the photon selection criteria.
In general, the results from both algorithms were found to be
very similar except for a somewhat stronger reducible
background rejection of the PYTHIA algorithm.
Therefore the numbers
presented in the following are from the PUCLUS jet finder.
For the photon-jet candidate, $\gamma_H$, it was
demanded that it involves only
one dominant neutral electromagnetic
shower compatible with originating from a single photon
and was not associated with any charged particle.
This jet should also not contain any neutral hadronic
activity and if it was accompanied by one or two further
photons, their total energy should not exceed 10\% of the
jet energy.
Finally we required that $\gamma_H$ has an energy transverse
to the beams, $E_T$, greater than 15 GeV, which is exploited to increase the
signal-to-background ratio. A variation of $E_T$
between 8 and 20 GeV verified that the value chosen optimizes in some
way $S/B$ and ensures a statistically significant signal event rate.
If one of these requirements failed the photon-jet candidate
was discarded and, if no other
candidate was found, the event was rejected from the analysis.

\vspace{0.3cm}
The resulting dijet invariant mass, $M_{jj}$, is clearly
dominated by the $Z$ boson at 91 GeV, with a total width at half-maximum
of approximately 4.5 GeV. This gives us confidence that the jet
clustering algorithms used isolates adequately the hadronic jets
and the $\gamma_H$ candidate might result from the Higgs decay.
To preserve good mass resolution for the final $Z \gamma$ system
the two hadronic jets were fitted to the constraint
$M_{jj} = M_Z$.

\vspace{0.3cm}
Remaining reducible background, mainly due to $e \nu W (\gamma)$,
\ee $(\gamma^{*}/Z) (\gamma)$ and radiative return events, 
was found to grow slowly with 
$M_{Z\gamma}$ having a broad maximum around 140 GeV. It contributes
at most 42\% of the irreducible background
in the Higgs mass region considered.
Fig.~4 shows the reconstructed $Z \gamma$ mass
for signal and summed background events after application of the jet finder
algorithm PUCLUS and the cuts mentioned above.
Higgs signals are visible at the assumed masses, with most abundant
Higgs production at 140 GeV. Once the Higgs mass is known
from searches in dominant decay channels, reliable BF(\hZg) estimates
are possible even for the worst case of $M_H$ = 120 GeV.

\vspace{0.3cm}
Selection efficiencies
for the $ \nu_e \bar{\nu_e} Z \gamma
\rightarrow \nu_e\bar{\nu_e}q\bar{q}\gamma$ signature,
the number of signal $(S)$ and irreducible background $(B)$ events
in the mass range $M_{Z\gamma}$ between 117 and 123 (137-143, 157-163) 
GeV, the significances
$S/\sqrt{B}$ and the statistical precisions $\sqrt{S+B}/S$ obtained on
$\sigma$(\ee $\rightarrow \nu_e \bar{\nu_e} H)\cdot
BF(H \rightarrow Z \gamma)\cdot BF(Z \rightarrow q\bar{q})$
are presented in Table~3.

\vspace{0.3cm}
Thus, application of the PYTHIA jet finder for Higgs event selection and
accounting for all $Z$ decays as well as the surviving reducible background
yields for the relative precisions on the \hZg branching fraction
$\Delta$BF(\hZg)/BF(\hZg) = 70\% (31\%, 57\%)
for $M_H$ = 120 (140, 160) GeV, after convolution with the
uncertainty of the
$\nu_e \bar{\nu_e} H$ production cross-section \cite{TDR} and
assuming 1 ab$^{-1}$ integrated luminosity
at $\sqrt{s}$ = 500 GeV.

\begin{center}
\begin{figure}[t] 
\begin{minipage}[t]{18cm}
\vspace{-1.0cm}
\epsfig{file=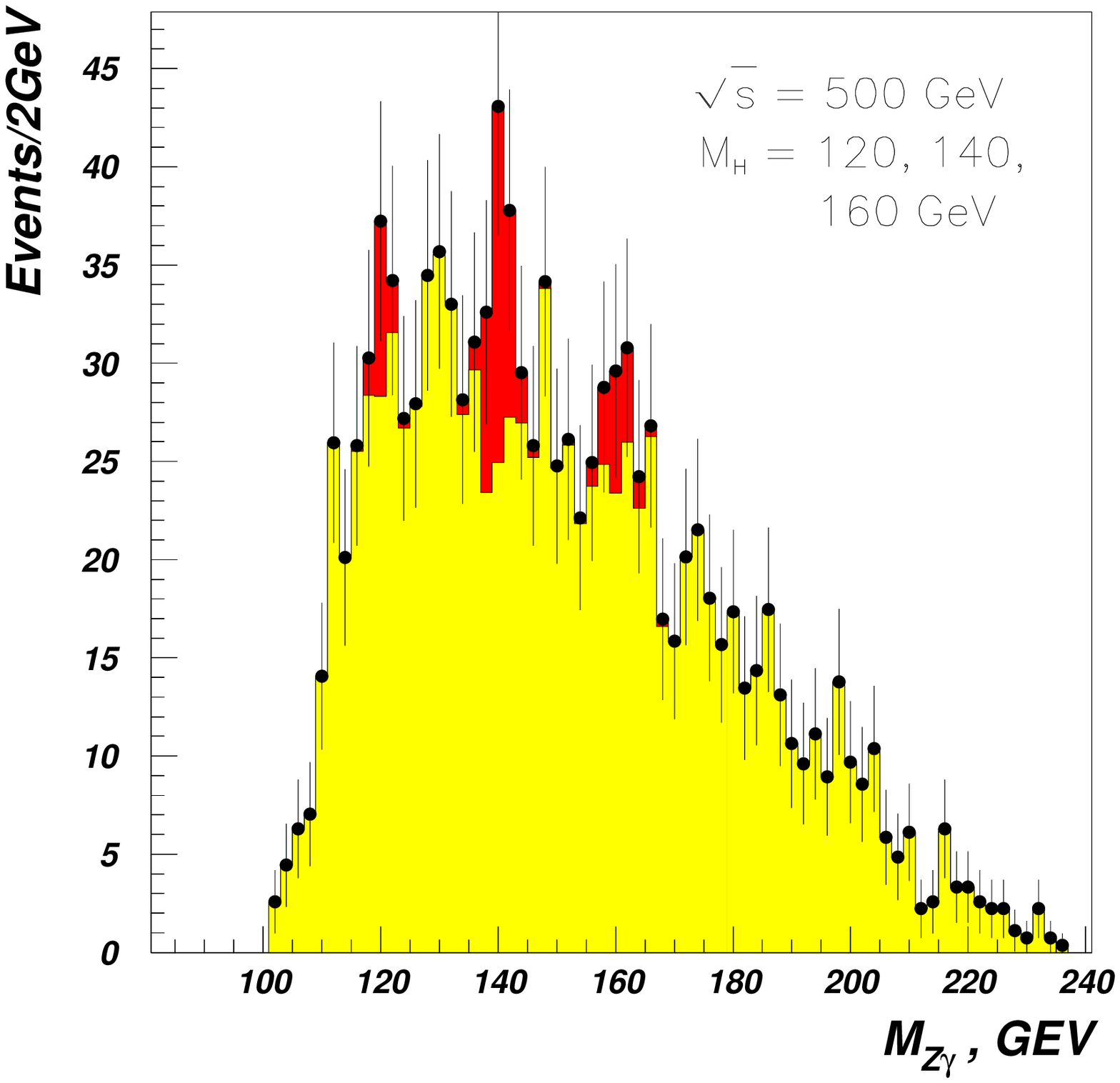,width=15cm}
\caption{ $M_{Z \gamma}$ invariant mass distributions for surviving
     signal and background events at $\sqrt{s}$ = 500 GeV,
     assuming 1 ab$^{-1}$ integrated luminosity.}
\vspace{0.5cm}
\end{minipage}
\end{figure}
\end{center} 

\begin{center}
\begin{tabular}{c|c|c|c} \hline \hline
               & \multicolumn{3}{c}{ Signal \hspace{3mm}(Background)}
 \\ \hline
               & $M_H$ = 120 GeV   &  140 GeV  &  160 GeV
 \\ \hline\hline
Selection efficiency (\%)   & 13.0 (0.27) & 13.9 (0.26) & 13.9 (0.25) \\ \hline
Number of events/1 ab$^{-1}$     & 12 (59) & 29 (57) & 14 (55) \\ \hline
Significance $S/\sqrt{B}$        & 1.56  & 3.84  & 1.89 \\ \hline
Precision $\sqrt{S+B}/S$         & 0.70  & 0.32  & 0.54 \\ \hline
\end{tabular}
\end{center}
 Table~3: Jet finder selection efficiencies for both signal and
background $\nu_e\bar{\nu_e}q\bar{q}\gamma$ signature, together with
significances and precisions of
$\sigma$(\ee $\rightarrow \nu_e \bar{\nu_e} H)\cdot
BF(H \rightarrow Z \gamma)\cdot BF(Z \rightarrow q\bar{q})$
for $M_H$ = 120, 140 and 160 GeV, at the \ee linear collider with
$\sqrt{s}$ = 500 GeV. A total integrated luminosity of 1 ab$^{-1}$
is assumed. \\

\subsection {\boldmath Event selection by means of 'Higgs-likeness'}

The results obtained by exploiting
consecutive cut or jet finder techniques
lead to small signal samples
accompanied by large backgound. Demanding further cuts beyond
the ones discussed so far does not improve $S/B$
significantly, but would instead reduce signal event rates
to a level precluding reasonable \hZg branching fraction measurements.
This is mainly because the irreducible background has similar
final state signature as the \hZg events and exceed this
sample by typically two or more orders of magnitude before any
selection procedure.

\vspace{0.3cm}
Such a situation calls for a more sophisticated selection approach
where also slight differences between signal and background events
are taken into account.
In this respect, a 'likelihood factor' was constructed giving a measure
of the probability that an event is part of the signal. For any
particular event, kinematical variables of the final state photon,
the $Z$ boson respectively the two jets and the missing
neutrino system
were combined into a global discriminant variable~$P_H$.
This quantity was constructed
from a variety of normalised variables based on large statistics
samples of simulated signal and background events.
The variables used account for possible kinematic
differences between the Higgs events (diagram 1 in Fig.~1)
with the isotropic \hZg decay
and the background (diagrams of Fig.~2). In particular,
the transverse momentum of the photon, its
cms scattering angle, the cosine of the polar angle of the $Z$ boson,
the photon polar and azimuthal decay angles in the Higgs rest frame,
the cms polar and azimuthal angles between the photon
and the $Z$, the cms photon energy, the collinearity
angle between the electron beam and the photon, the coplanarity
angles of the beam, the photon and the $Z$ boson as well
as the beam, the Higgs and the photon in the Higgs rest frame,
the transverse masses of the photon and the missing system as well as
the Higgs and the missing system were considered.
In events where more than one photon candidate exists
(about 48 \% of the cases) the photon with largest energy was
selected as the Higgs decay candidate.
For each event which passes the principal cuts, the event
and jet quality cuts (see sect.~3.1) and the fit constraint
$M_{jj} = M_Z$, signal and background
probabilities were then calculated, and by multiplication
of all signal probabilities the sensitivity for an event to be a
Higgs candidate was maximised.
The quantity so obtained is constraint to lie in the region [0;1].
Background events are preferably distributed at low $P_H$ values while
for Higgs signal events $P_H$ is close to unity.
Since several variables included in the analysis vary with
the Higgs mass, signal probabilities were individually determined
for $M_H$ = 120, 140 and 160 GeV. Therefore, the
'Higgs-likeness' exists for each Higgs mass considered.
Fig.~5 shows, as an example,
$P_H$ for $M_H$ = 120 GeV signal and the sum of signal
and background events.
Similar distributions were obtaind for $M_H$ = 140 and 160 GeV.
Finally, only events were retained if the energy of the
Higgs photon candidate, $\gamma_H$, was greater than 20 GeV
with the transverse component $E_T >$ 15 GeV.
Reducible background was found to arise from $W W (\gamma)$,
$e \nu W (\gamma)$ and radiative return events, with rates of at most 32\% of the
irreducible background at the 160 GeV Higgs mass.

\vspace{0.3cm}
Fig.~6 shows the $Z \gamma$ invariant mass spectra
for the luminosity adjusted $\nu_e \bar{\nu_e} Z \gamma$
signal and background events surviving the cut $P_H > 0.98$,
for $M_H$ = 120, 140 and 160 GeV. In all cases, convincing
Higgs signals are evident on non-negligible backgrounds.
Compared to the selection procedures discussed in the previous sections
best Higgs signal significancies were obtained. 
In particular, the excess of events at 140 GeV is very encouraging. 
But also the somwhat
degraded event rates at $M_H$ = 120 and 160 GeV allow for reliable
\hZg branching fraction estimates.
Variation of the discriminant
variable $P_H$ between 0.82 and 0.98 does not improve $S/B$,
but would rather lower the signal-to-background ratio.

\vspace{0.3cm}
Selection efficiencies
for the $\nu_e \bar{\nu_e} Z \gamma
\rightarrow \nu_e\bar{\nu_e}q\bar{q}\gamma$ signature,
the number of signal $(S)$ and irreducible background $(B)$ events
in the mass range
$M_{Z\gamma}$ between 117 and 123 (137-143, 157-163) GeV,
the significances
$S/\sqrt{B}$ and the statistical precisions $\sqrt{S+B}/S$ obtained on
$\sigma$(\ee $\rightarrow \nu_e \bar{\nu_e} H)\cdot
BF(H \rightarrow Z \gamma)\cdot BF(Z \rightarrow q\bar{q})$
are presented in Table~4.

\begin{center}
\begin{tabular}{c|c|c|c} \hline \hline
               & \multicolumn{3}{c}{ Signal \hspace{3mm}(Background)}
 \\ \hline
               & $M_H$ = 120 GeV   &  140 GeV  &  160 GeV
 \\ \hline\hline
Selection efficiency (\%)   & 17.4 (0.23) & 13.9 (0.20) & 15.9 (0.16) \\ \hline
Number of events/1 ab$^{-1}$     & 16 (51) & 29 (44) & 16 (41) \\ \hline
Significance $S/\sqrt{B}$        & 2.24  & 4.37  & 2.50 \\ \hline
Precision $\sqrt{S+B}/S$         & 0.51  & 0.29  & 0.47 \\ \hline
\end{tabular}
\end{center}
 Table~4: "Higgs-likeness" selection efficiencies
for both signal and
background $\nu_e\bar{\nu_e}q\bar{q}\gamma$ signature, together with
significances and precisions of
$\sigma$(\ee $\rightarrow \nu_e \bar{\nu_e} H)\cdot
BF(H \rightarrow Z \gamma)\cdot BF(Z \rightarrow q\bar{q})$
for $M_H$ = 120, 140 and 160 GeV, at the \ee linear collider with
$\sqrt{s}$ = 500 GeV. A total integrated luminosity of 1 ab$^{-1}$
is assumed. \\

Accounting for all Z decays and the surviving reducible background
the relative precisions on the \hZg branching fraction
are then deduced after convolution with the uncertainty of the
inclusive $\nu_e \bar{\nu_e} H$ production rates \cite{TDR}:
$\Delta$BF(\hZg)/BF(\hZg) = 48\% (27\%, 44\%)
for $M_H$ = 120 (140, 160) GeV, assuming
1 ab$^{-1}$ integrated luminosity
at $\sqrt{s}$ = 500 GeV.

\begin{center}
\begin{figure}[t]
\vspace{-2.0cm}
\begin{minipage}[b]{14cm}
\epsfig{file=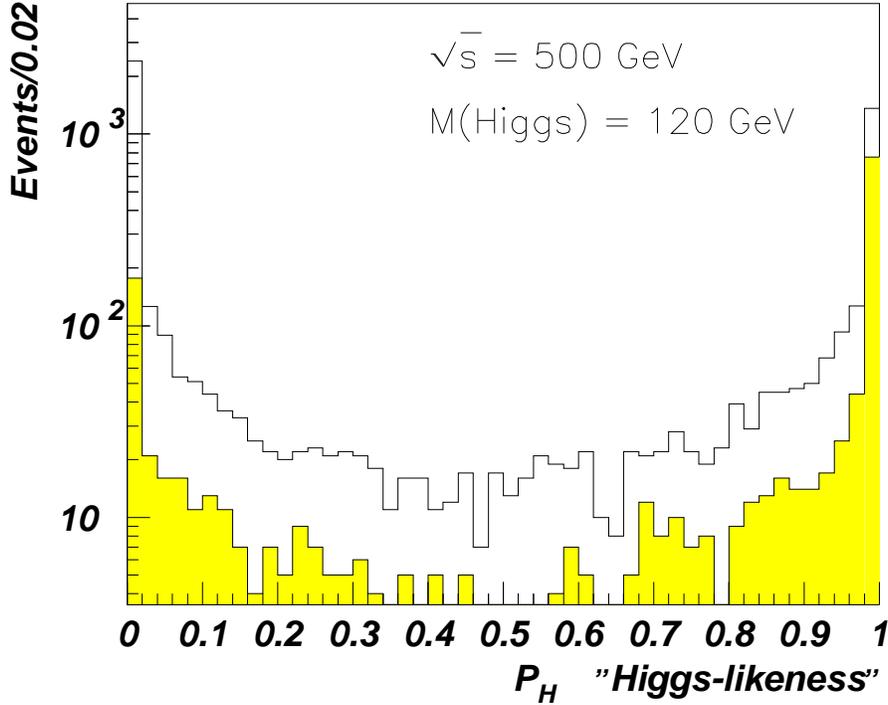,width=14cm}
\caption{Distribution of the discriminant variable $P_H$ for
   \ee $\rightarrow \nu_e \bar{\nu_e} H \rightarrow \nu_e \bar{\nu_e} Z \gamma$
   signal events (shaded histogram) and the sum of signal and
   background contributions.}
\vspace{0.5cm}
\end{minipage}
\end{figure}
\end{center} 

\begin{center}
\begin{figure}
\vspace{-0.3cm}
\includegraphics[width=0.80\textwidth,height=.90\textheight]
  {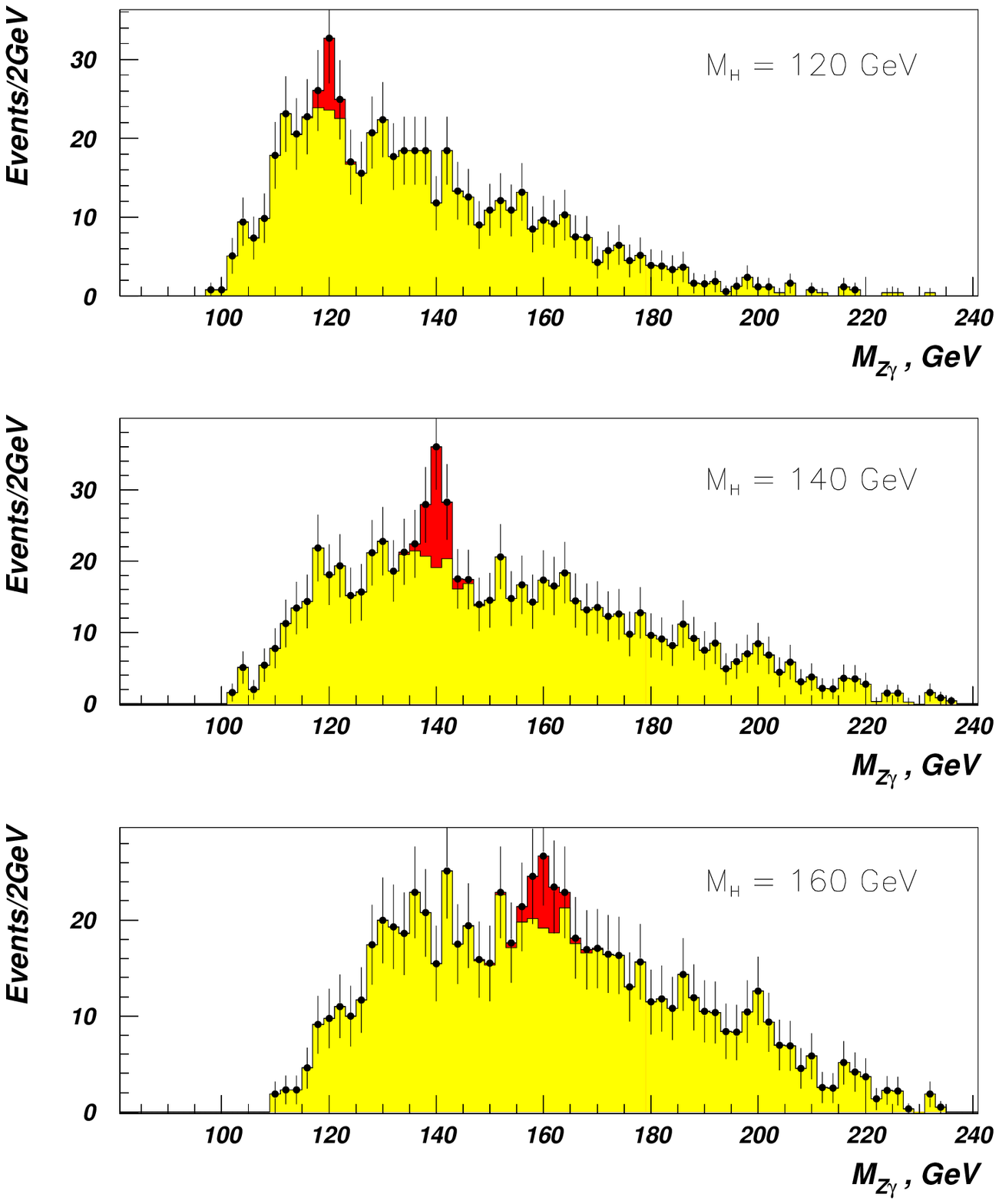}
\vspace{-0.3cm}
\caption{ $M_{Z \gamma}$ invariant mass distributions for surviving
     signal and background events at $\sqrt{s}$ = 500 GeV, for
     $M_H$ = 120, 140 and 160 GeV and 1 ab$^{-1}$ integrated luminosity.}
\end{figure}
\end{center}

\subsection {\boldmath Discussion of the results}

Relatively independent on the selection technique exploited
we would register typically 15 to 30 \hZg events in reaction (3)
and three or more times background events in the window $M_H \pm 3$ GeV
at $\sqrt{s}$ = 500 GeV for an integrated luminosity of 1 ab$^{-1}$.
A comparison of the selection procedures applied favors the
'Higgs-likeness' method, in particular for the 120 GeV Higgs mass case.
The results attainable for the relative uncertainties of the
\hZg branching fraction are 48\% (27\%, 44\%)
for $M_H$ = 120 (140, 160) GeV.
We regard these numbers as rather encouraging, especially considering
the initial value of the S/B rates. Although a complete optimization
for neither selection technique has not been achieved, we are
confident that room for improved \hZg branching fraction measurements is limited,
mainly due to the presence of overwhelming irreducible background 
with final state signature
similar to the signal events. Improvements may rely
on the size of the Higgs window not yet
adjusted to obtain optimized numbers of signal to background
events for a narrow Gaussian resonance whose observed width is dominated
by instrumental effects. Also Higgs-strahlung events which 
exist in the data sample at 500 GeV would provide an independent
\hZg branching fraction, based on however different selection procedures.
Although a less accurate measurement is anticipated (due to larger
\ee $\rightarrow Z Z \gamma$ irreducible background and more complicated
analyses when both Z bosons decay hadronically), but in combination with
the $W W$ fusion measurement an improved $\Delta$B(\hZg)/B(\hZg)
of about 10\% can be expected.
Furthermore, inclusion of
leptonic $Z$ boson decays and $Z Z$ fusion events of reaction (4),
so far neglected, would slightly improve the results.
However, the increase of the signal event sample of about 10 to 20\%
would be partially compensated by more background which scales
in approximatelly the same way.

\vspace{3mm}
An alternative approach relies on studying the Higgs-strahlung
process
$ e^+e^- \rightarrow H Z  \rightarrow Z Z \gamma$ at optimezed lower
energies, e.g. at $\sqrt{s}$ = 300 GeV. Here, approximately 95\%
of all Z decays are useful and for ${\cal L}$ = 1 ab$^{-1}$
about 200 $H Z \rightarrow Z Z \gamma $ events are expected.
Assuming similar selection efficiencies
as found for the $W W$ fusion process, the final \hZg sample
would consist of some 50 events,
of which approximately 50\% are 4-jet events with a prompt
photon. A topology of four jets with an accompaning photon
is not only produced by
\hZg $\rightarrow q\bar{q}\gamma$ decays, but also via
fragmentation of quarks and gluons of continuum $Z Z$ production.
Refering also to the huge
irreducible $ e^+e^- \rightarrow Z Z \gamma$ background
and the large cross-section process
$ e^+e^- \rightarrow W W (\gamma)$ with initial and final state
photon radiation, detailed analyses are needed
and preliminary results indicate less precision on BF(\hZg)
\cite{our,Cubitt}.

\vspace{3mm}
Linear \ee colliders offer the possibility for longitudinal
polarised electron and positron beams, with varying polarisation degrees
in right-handed or left-handed modes.
Higgs boson production rates in both processes (2) and (3)
depends strongly on the polarisation degree and the
helicity of the incoming particles.
For any given process, ratios of the cross-section
for given electron and positron beam polarisations divided
by the cross-section for unpolarised beams, denoted as $R$, are presented
in Table 2 of \cite{shanidze}
for different beam polarisations. The Higgs event rates 
in processes (2) and (3)
are enhanced most for left-handed $e^-$ colliding with right-handed
$e^+$ with as large a degree of polarisation as possible.
For the feasible though ambitious case of collisions between an
$e^-$ beam with polarisation $P_-$ = -0.8 and an $e^+$ beam with
polarisation $P_+$ = +0.6, $R$ = 1.77 and 2.88 for the
Higgs-strahlung and $W W$ fusion processses, respectively.
However, the dominant irreducible background in both processes
scales in approximately the same way with beam
polarisations as the signal processes \cite{shanidze},
the precision of the \hZg branching fraction
improves by only a factor $\sqrt{R}$. Under such circumstances,
the relative uncertaities on BF(\hZg) is lowered to 28\% (16\%, 26\%)
for $M_H$ = 120 (140, 160) GeV, if the "Higgs-likeness" selection
technique is exploited. However, it should be noted that other
physics processes will demand different beam polarisations and the
assumption of using the full luminosity with the desired
beam polarisation for this particular measurement gives some lower
bound to the attainable precision.

\vspace{3mm}
Since the signal-to-background ratio is expected to be less than unity,
it should be emphasized that large continuum $Z\gamma$ production 
and copious $\pi^0$ background events must be rejected by excellent
geometrical resolution and stringent isolation criteria combined with
excellent electromagnetic and hadronic energy resolution and
hermiticity. A worse resolution would flatten the marginal signal events
over large $Z\gamma$ background, thus degrading the visibility of the signal.
Systematic uncertainties due to detector effects
such as photon detection effeciency, energy scales and
resolutions are believed to be small and can be estimated from
comparison of data with well understood processes,
such as $ e^+e^- \rightarrow \gamma \gamma$, Compton scattering,
Bhabha, $Z Z $ and $W W$ events.
Systematic uncertainties on the integrated luminosity
are expected to be below 0.5\%, and statistical uncertainties
due to final simulation sample sizes should be kept below few percent.
Simulations of the Standard Model background channels are expected
to yield most of the sytematic uncertainties. The use of different
event generators would keep this uncertainty under control and
agreement between them within few percent is expected.
Taking all these effects together and accounting for a  
precise measurement of the inclusive Higgs cross section of about 4\%
or less \cite{TDR},
it appears that the error on BF(\hZg) will be dominated by
the statistical uncertainty.

\vspace{3mm}
The effect of overlap of $\gamma \gamma \rightarrow hadrons$
to $\nu_e \bar{\nu_e} H$ events has also been studied. The
$\gamma \gamma$ events due to photons radiated in the
electro-magnetic interactions of the colliding beams have been generated
by the GUINEA PIG program \cite{GUIN} with a rate modelled in
\cite{Schuler}. PYTHIA \cite{PYTHIA} has been used to
generate the hadrons. The latest version of SIMDET
\cite{SIMDET} overlays the $\gamma \gamma$ events
to $e^+e^- \rightarrow \nu_e \bar{\nu_e} H$ events,
and all final state particles are then reconstructed. Without special
care to isolate the particles from $\gamma \gamma$ interactions,
we found that the Higgs events were recognized without
notable loss or distortions after passing any of the selection
procedure.

\section{Conclusions}

We have examined the prospects at a future linear \ee collider
of measuring the branching fraction of a Standard Model-like Higgs boson
into the $Z$ boson and a photon, BF(\hZg). Higgs boson masses of
120, 140 and 160 GeV and in integrated luminosity of 
${\cal L}$ = 1 ab$^{-1}$
at $\sqrt{s}$ = 500 GeV were assumed. In order to estimate
the precision on BF(\hZg) which can be attained, all expected
background processes were included in the analysis, and acceptances
and resolutions of a linear collider detector were taken into account.
In particular, by simulating the 2-to-4 particle reactions
\ee $\rightarrow \nu_e \bar{\nu_e} Z \gamma$, in which the signal
reaction $e^+e^- \rightarrow \nu_e \bar{\nu_e} H$ is embedded,
the complete irreducible background
has been taken into account. Only $ Z\rightarrow$ \qq decays were included
so far.

Since reactions like \ee $\rightarrow \nu_e \bar{\nu_e} Z \gamma,
W W (\gamma)$, $e \nu W (\gamma)$,
$e^+ e^- (\gamma^*/Z)$ and radiative return $q\bar{q}\gamma$ events
also constitute potentially serious background
sources for the $e^+e^- \rightarrow \nu_e \bar{\nu_e} H$ signal,
different selection techniques (consecutive cuts,
jet finders, 'Higgs-likeness') were applied and have been shown
to result in tolerable background levels and 
Higgs detection in the rare \hZg decay.
As the favored selection procedure the 'Higgs-likeness'
technique has been found, with 15 to 30 identified signal events,
comparable to the other methods,
but with lowest total background.

For unpolarized beams, the expected relative precision for the \hZg
branching fraction was found to be 48\% (27\%, 44\%)
for $M_H$ = 120 (140, 160) GeV, after accounting for all $Z$ decays
and convolution with the uncertainty on the
inclusive $W W$ fusion Higgs boson cross-section.

For $e^-$ beam polarisation of -0.8 and $e^+$ beam polarisation
of +0.6, the $W W$ fusion cross-section 
$\sigma$(\ee $\rightarrow \nu_e \bar{\nu_e} H)$
is significantly
enhanced, so improving substantially the
precision on BF(\hZg) to 28$\%$ (17$\%$, 26$\%$), even taking into
account the fact that the dominant irreducible background
scales in the same way. With these uncertainties it should be possible
to deduce a relative precision for the \hZg partial width of
$\frac{\Delta\Gamma(H \rightarrow Z \gamma)}{\Gamma(H \rightarrow Z\gamma)} \simeq 29\% (17\%, 27\%)$,
if an uncertainty of 5\% for the total Higgs width \cite{TDR} is included.
This in turn allows to expect a relative precision for the
$H Z \gamma$ coupling of 15\% (9\%, 14\%)
for $M_H$ = 120 (140, 160) GeV.

Overlap of $\gamma \gamma \rightarrow hadrons$ to
$\nu_e \bar{\nu_e} H$ events due to photons radiated in the
electro-magnetic interactions of the colliding beams
would not alter the uncertainties accessible.

The results presented also suggest that the $W W$ fusion reaction
$e^+e^- \rightarrow \nu_e \bar{\nu_e} H$ at 500 GeV
would be superior in \hZg branching fraction measurements
to the Higgs-strahlung process
\ee $\rightarrow H Z $ at lower energies,
e.g. at $\sqrt{s}$ = 300 GeV \cite{our,Cubitt},
in particular if polarised beams are taken into account.
However, detailed analyses are needed for the latter process
to establish present indications.

For Higgs masses significantly above 160 GeV, it will be difficult 
to determine the $H Z \gamma$ coupling with valuable precision
since the \hZg branching fraction is too
small to be accurately measured.

\vspace{3mm}
In summary, our results confirms the unique ability
of a linear \ee collider to access and reliably measure fundamental
Higgs boson parameters in the presence of initially overwhelming
background with final state signature similar to the signal
events.

\vspace{0.5cm}
\section*{Acknowledgments}
We would like to thank K.Desch and E.Boos for useful discussions.
The work of M.D. was partially supported by RFBR grant 01-02-1670
and INTAS grants 00-00313, 00-00679.



\begin{thebibliography}{99}

\bibitem{LC1} e.g. E. Accomando et al., Phys. Rep. 299 (1998) 1. 
%
\bibitem{couplings} for a review see e.g. K. Desch and M. Battaglia,
 in Physics and experiments with future linear \ee colliders,
 Proceedings of the 5th International Linear Collider Workshop,
 Batavia, IL, USA, 2000. 
%
\bibitem{tricoupl} G. Gounaris, D. Schildknecht and F. Renard, Phys. Lett.
 B 83 (1979) 191 \\
 and B 89 (1980) 437; \\
 V. Barger and T. Han, Mod. Phys. Lett. A 5 (1990) 667; \\
 A. Djouadi, H.E. Haber and P.M. Zerwas, Phys. Lett. B 375 (1996) 203; \\
 V. Ilyin et al., KEK CP-030; \\
 F. Boudjema and E. Chopin, Report ENSLAPP-A534/95; \\
 D.J. Miller and S. Moretti, Eur. Phys. J. C 13 (2000) 459; \\
 C. Castanier et al., LC-PHSM-2000-061. 
%
\bibitem{loop} J.~Ellis, M.K.~Gaillard and D.V.~Nanopoulos, Nucl. Phys. B
106 (1976) 292; \\
A.I.~Vainshtein et al., Sov. J. Nucl. Phys. 30 (1979) 711; \\
R.N.~Cahn, M.S.~Chanowitz and N.~Fleishon, Phys. Lett. B 82 (1979) 113; \\
L.~Bergstrom and G. Hulth, Nucl. Phys. B 259 (1985) 137. 
%
\bibitem{Hgg} M. Battaglia, Proceedings of the Wordwide Study on
 Physics and Experiments with Future Linear \ee Colliders,
 Sitges, Barcelona, Spain, April 28 - May 5, 1999. 
%
\bibitem{HggLHC} H.M. Georgi et al., Phys. Rev. Lett. 40 (1978) 692; \\
A. Djouadi, M. Spira and P.M. Zerwas, Phys. Lett. B 264 (1991) 440; \\
S. Dawson, Nucl. Phys. B 359 (1999) 283; \\
M. Spira et al., Nucl. Phys. B 453 (1995) 17. 
%
\bibitem{shanidze} E. Boos et al., Eur. Phys. J. C 19 (2001) 455. 
%
\bibitem{gagaLHC} ATLAS Collaboration, "ATLAS detector and physics
performance, technical design report", vol. 2, report  CERN/LHCC 99-15
ATLAS-TDR-15; \\
CMS Collaboration, "CMS: The electromagnetic calorimeter, technical
design report", report CERN/LHCC 97-33, CMS-TDR-4. 
%
\bibitem{gammagamma} G. Jikia and S. S\"oldner-Rembold, 
Nucl. Phys. Proc. Suppl. 82 (2000) 373; \\
G. Jikia and S. S\"oldner-Rembold, LC-PHSM-2001-060; \\
M. Melles, W.J. Stirling and V.A. Khoze, Phys. Rev. D61 (2000) 54015. 
%
\bibitem{La} e.g. G. Degrassi, hep-ph/0102137; \\
J. Erler, hep-ph/0102143; \\
 D. Abbaneo et al. [LEP Electroweak  Working Group] and A. Chou et al.
 [SLD Heavy Flavour and Electroweak Groups], LEPEWWG/2002-01 (May 2002) \\
 and additional updates at {\tt http://lepewwg.web.cer.ch/LEPEWWG/}. 
%
\bibitem{LEP} ALEPH Collaboration, R. Barate et al., Phys. Lett. B
 495 (2000), 1; \\
 DELPHI Collaboration, P. Abreu et al., Phys. Lett. B 499 (2001) 23; \\
 L3 Collaboration, M Acciari et al., Phys. Lett. B 508 (2001) 225; \\
 OPAL Collaboration, G. Abbiendi et al., Phys. Lett. B 499 (2001) 38; \\
 The LEP Working Group for Higgs Boson Searches, hep-ex/0107029 \\
 and LHWG Note 2002-01 (July 2002) \\
 and additional updates at
 {\tt http://lephiggs.web.cern.ch/LEPHIGGS/www/Welcome.html}. 
%
\bibitem{CompHEP} E.E.Boos et al., INP MSU 94-36/358 and SNUTP-94-116,
 hep-ph/9503280; \\
 P. Baikov et al., Proc. of the Xth Int. Workshop on High Energy
 Physics and Quantum Field Theory, QFTHEP-95, ed. by B. Levtchenko
 and V. Savrin, Moscow, 1995, p.101; \\
 A. Pukhov et. al., CompHEP user's manual, v.3.3, INP MSU 98-41.542 and
 hep-ph/9908288. 
%
\bibitem{bms} D. Schulte, private communication; \\ 
 T. Ohl, IKDA  96/13-rev., July 1996 and hep-ph/9607454-rev. 
%
\bibitem{PDG} D.E. Groom et al., Eur. Phys. J. C 15 (2000) 1. 
%
\bibitem{interface} A.S. Belyaev et al., hep-ph/0101232. 
%
\bibitem{HDECAY} A. Djouadi, J. Kalinowski and M. Spira,
 Comput. Phys. Commun. 108 (1998) 56. 
%
\bibitem{PYTHIA} T. Sj\"{o}strand, Comput. Phys. Commun. 82 (1994) 74; \\
 T. Sj\"{o}strand, hep-ph/0001032; \\
 T. Sj\"{o}strand et al., Comput. Phys. Commun. 135 (2001) 238. 
%
\bibitem{our} M. Dubinin, H.J. Schreiber and A. Vologdin, in
  preparation. 
%
\bibitem{SIMDET} M. Pohl and H.J. Schreiber, DESY 02-061, May 2002,
 hep-ex/0206009, \\ LC-DET-2002-005. 
%
\bibitem{TDR} TESLA Technical Design Report, Part IV, A Detector
 for TESLA, DESY 2001-011, ECFA 2001-209, hep-ph/0106315. 
%
\bibitem{Durham} S. Bethke et al., Nucl. Phys. B 370 (1992) 310. 
%
\bibitem{Cubitt} T. Cubitt, DESY Zeuthen summer student write-up,
October 2001. 
%
\bibitem{GUIN} D. Schulte, TESLA note 97-08. 
%
\bibitem{Schuler} G. A. Schuler and T. Sj\"{o}strand, CERN-TH/96-119. \\

\end{thebibliography}
\end {document}